\documentclass[pra, showpacs,twocolumn,eqsecnum,superscriptaddress]{revtex4}
\usepackage{makeidx}
\usepackage{eurosym}
\usepackage{amssymb}
\usepackage{amsfonts}
\usepackage{amsmath}
\usepackage{setspace}
\usepackage[english]{babel}
\usepackage{graphicx}
\usepackage[applemac]{inputenc}
\usepackage{color}
\usepackage{ifthen}
\usepackage{float}
\usepackage{verbatim}
\usepackage{subfigure}

\begin{document}

\title{Quantum phase transitions of light in a dissipative Dicke-Bose-Hubbard model}

\author{Ren-Cun Wu}
\affiliation{Institute of Theoretical Physics, Lanzhou University, Lanzhou $730000$, China}
\author{Lei Tan}
\email{tanlei@lzu.edu.cn}
\affiliation{Institute of Theoretical Physics, Lanzhou University, Lanzhou $730000$, China}

\author{Wen-Xuan Zhang}
\affiliation{Institute of Theoretical Physics, Lanzhou University, Lanzhou $730000$, China}
\author{Wu-Ming Liu}
\affiliation{Beijing National Laboratory for Condensed Matter Physics, Institute of Physics, Chinese Academy of Sciences, Beijing 100190,China}
\date{\today}
\begin{abstract}
The impacts that the environment has on the quantum phase transition of light in the Dicke-Bose-Hubbard model are investigated. Based on the quasibosonic approach, mean field theory and the perturbation theory, the formulation of the Hamiltonian, the eigenenergies and the superfluid order parameter are obtained analytically. Compared with the ideal cases, the order parameter of the system evolves with time as the photons naturally decay in their environment. When the system starts with the superfluid state, the dissipation makes the photons tend to localize, and a greater hopping energy of photon is required to restore the long-range phase coherence of the localized state of the system. Furthermore, the Mott lobes disappears and the system tends to be classical with the number of atoms increasing; however, the atomic number is far lower than that expected under ideal circumstances. Therefore, our theoretical results offer valuable insight into the quantum phase transition of a dissipative system.
\end{abstract}

\pacs{42.50.Nn, 42.50.Pq, 05.70.Fh} 
\maketitle


\section{Introduction}
Quantum simulation has become a research frontier and an indispensable tool in quantum information science\cite{M,Hur,Noh},  it's remarkable development in experiment realization has led to incredible advances in the field of quantum optics and atomic physics\cite{Michael, Birnbaum,Mabuchi,Raimond,Aoki,Srinivasan}. Among the recent developments, the system of coupled cavity arrays embedded with cold atoms has been intensively investigated as a platform to realize and simulate quantum many body phenomena because of its extremely high tunability, individual addressability and flexibility in its geometric design\cite{Toyoda,Underwood,Hartmann,Greentree,Angelakis}. A wide range of condensed matter system have been theoretically investigated and  many proposals for probing them have been proposed including the quantum phase transition\cite{Koch,Brandao}, spin glasses\cite{Strack,Li}, photon crystals\cite{Jin}, the emergence of gauge fields\cite{Umucal1lar}, the quantum Hall effects\cite{Carusotto}, the pfaffian-like topological state\cite{Andrew} and the supersolid \cite{Bujnowski,Guo}.

The simplest physical model of light-matter coupling in a coupled cavity array system is the Jaynes-Cummings Hubbard model, which presents an array of optical cavities that each contain a single two-level atom(TLA) in the photon-blockade regime\cite{Greentree,Angelakis}. A modified Jaynes-Cummings Hubbard model based on each cavity embedded a three-level atom has been proposed recently; this model circumvents the drawbacks of the excited-state spontaneous emission and provides a tunable extension of two-polariton bound states of the standard Jaynes-Cummings Hubbard \cite{Maggitti,Minar}. As the number of atoms in each cavity increases, the collective effects due to atomic interactions among themselves give rise to intriguing many-body phenomena. In quantum optics the Dicke model is a paradigm of collective behavior\cite{Dicke} that describes the interaction of ensembles of TLAs that are collectively coupled to the single mode of radiation of a cavity\cite{Badshah}. Numerous investigations of interesting physical effects and their experiment realization\cite{Chitra}, such as the super-radiation phase\cite{Mlynek,Hepp,Baumann}, the superradiant Mott Insulator\cite{Klinder} and the dynamical phase transition\cite{Kebler}, are discussed. Thus, a Dicke-Bose-Hubbard (DBH) model that includes more than one identical coupled cavities and  $N$ identical TLAs in each cavity has been conducted to study the quantum phase transitions of light without considering the counter-rotating terms\cite{Lei}. The transfer of excitations under a large range of operative conditions is also demonstrated and explored by tuning the controlling parameters in the DBH model\cite{Badshah}. Both the emergence of a polaritonic glassy phase\cite{Rossini} and the quantum phase transitions from the superfluid to the Bose-glass and the Mott-insulator states\cite{Na} have also been studied. Most recently, the localization-delocalization quantum phase transition of photons of the DBH model including counter-rotating terms has been presented\cite{Lu}. The model shows that under the influence of the counter-rotating terms, the Mott lobes are fully suppressed.

As is well known, a realistic quantum optical system can rarely be isolated from its surrounding completely, particularly in an experiment; rather it is usually coupled to its external environment with an infinite number of degrees of freedom. To date, an investigation of quantum phase transition of photons in an dissipative DBH model is still lacking. To treat the interplay between the coupled cavity arrays and its environment in a more general setting, we developed a quasi-bosonic approach to describe the quantum phase transition and photon transport in an open quantum optical systems \cite{Liu,Tan}. Without the requirement of considering the finite environment's degrees of freedom, the quasibosonic method is a great concept, that has a computational advantage. In the present paper, we use the quasibosonic approach to obtain an effective Hamiltonian of the dissipative DBH model. The coordinates of a bath can be eliminated and the system can be considered an ensemble of quasi-bosons in less time than its decay rate. Next, the eigenenergies and the superfluid order parameter of the system are also derived analytically for two TLAs on resonance, and we numerically demonstrate the phase diagram of an arbitrary number of TLAs. The theoretical analysis presented here will be an essential reference for future experiments to explore the quantum effects for multi-atom system.

The paper is organized as follows. In section \uppercase\expandafter{\romannumeral2}, the dissipative DBH model is introduced based on the quasibosonic approach. Section \uppercase\expandafter{\romannumeral3} is devoted to deriving the eigenvalues and eigenstates for two atoms in each cavity. The analytical solutions of the superfluid order parameter for dressed states are given, and the properties of quantum phase transition are discussed in section \uppercase\expandafter{\romannumeral4}. The extension to
an arbitrary number of TLAs is also given in this section. Section \uppercase\expandafter{\romannumeral5} gives the conclusion.

\section{The dissipative Dicke-Bose-Hubbard model}
\begin{figure}[ht]
\includegraphics[width=6cm,height=6cm]{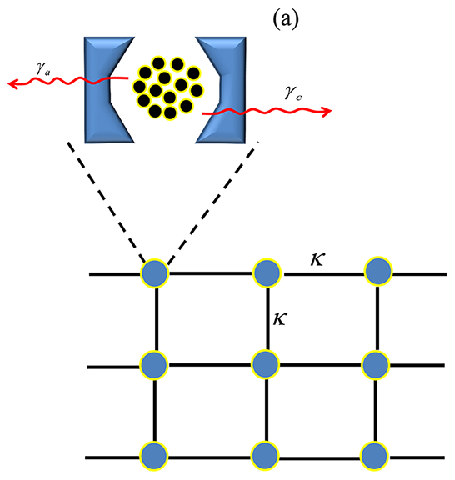}
\includegraphics[width=6cm,height=6cm]{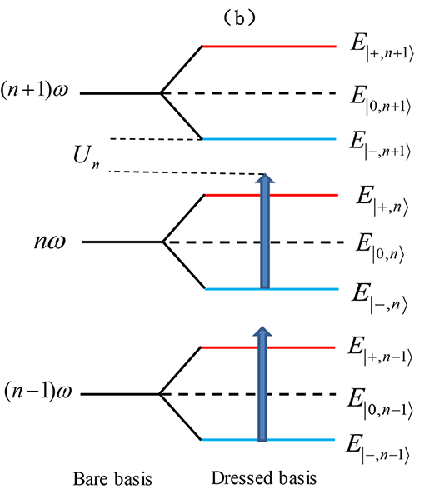}
\caption{(Color online) (a) A schematic of a two-dimensional coupled cavity array set up. Each cavity contains $N$ identical two-level atoms which couple resonantly to the cavity field. The yellow solid lines represent the interaction with the environment. Red wavy arrows indicate the cavity $(\gamma_c)$ and atom decay $(\gamma_a)$. (b) Energy level diagram of the cavity-atom system on $n$th site. The system is on resonance, i.e., $\tilde{\omega}_{c}=\tilde{\omega}_{a}=\tilde{\omega}$. The on-site repulsion $U_{n}$ can be defined as $U_{n}=E_{|-,n+1\rangle}-E_{|-,n\rangle}-\omega$, which impedes the absorption for the next photon.}
\end{figure}
The system considered is depicted in Fig. $1$. The Hamiltonian of the DBH model considering the coupling to its environment is give by (with $\hbar=1$)\cite{Lei}
\begin{eqnarray}
H&=&\sum_{i}H^{DM}_{i}-\kappa\sum_{ij}a^{\dag}_{i}a_{j}-\mu\sum_{i}N_{i}+H_{R}\nonumber\\
H^{DM}_{i}&=&\omega_{a}J^{+}_{i}J^{-}_{i}+\omega_{c}a^{\dag}_{i}a_{i}+\beta(a_{i}J^{+}_{i}+a^{\dag}_{i}J^{-}_{i})\nonumber\\
H_{R}&=&\sum_{i}\sum_{\omega_{k}}\omega_{k}r_{k}^{\dag}r_{k}+H_{CR}+H_{AR} \\
H_{CR}&=&\sum_{i}\sum_{\omega_{k}}[\eta_{c}(\omega_{k}) r_{k}^{\dag}a+h.c] \nonumber\\
H_{AR}&=&\sum_{i}\sum_{\omega_{k}}[\eta_{a}(\omega_{k}) r_{k}^{\dag}J^{-}+h.c] \nonumber
\end{eqnarray}
where the indices $i$ and $j$ are individual cavities range over all sites. $a^{\dag}_{i}$ and $a_{i}$ are the photon creation and annihilation operators, respectively. $J^{\pm}_{i}=\sum_{j}\delta^{\pm}_{j}$ are the atomic collective raising and lowering angular momentum operators and the total number of excitations is $N_{i}=a^{\dag}_{i}a_{i}+J^{+}_{i}J^{-}_{i}$. The transition energy of TLA is $\omega_{a}$ and $\omega_{c}$ is the frequency of the cavity field. All the atoms couple to cavities with the same coupling $\beta$\cite{Buzek}. We assume that the hopping energy of photons $\kappa_{ij}=\kappa$ between sites $i$ and $j$ and the chemical potential in the grand canonical ensemble $\mu_{i}=\mu$ are the same for all cavities. The coupling Hamiltonian of the system with the environment and the Hamiltonian of the environment are described as $H_{R}$. $\omega_{k}$ is a model of bath, and $r_{k}^{\dag}(r_{k})$ is the creation and annihilation operators of environment in the $k$th model. $H_{CR}$ is the interaction of cavity with environment. The interaction of the atoms with the environment is denoted as $H_{AR}$.

Considering the influence of the environment, the decoherence of every cavity and the two-level atom would result in the incoherent or dissipative propagation of the incident photon, thus  nonequilibrium dynamics for the open quantum many-body system will arise.
In general, simulations of nonequilibrium many-body effects for a finite freedom of the system can be performed using the master equation and the mean-field decoupling approximation\cite{Nissen,J,Gerace,Aron}. However it is a formidable task to solve a fairly large parameter space because of the infinite freedom of the environment. To address this problem, our group proposed a quasibosonic approach to eliminated the infinite freedom of the environment, in which the operators of the environment can be treated as a c-number and then the dissipative system can be solved easily\cite{Liu,Tan}. One can obtain an effective Hamiltonian for the system based on the quasibosonic approach.
\begin{eqnarray}
H &=& \sum_{i}H^{DM}_{i}-\kappa\sum_{ij}\tilde{a}^{\dag}_{i}\tilde{a}_{j}-\mu\sum_{i}\tilde{N_{i}}\nonumber\\
H^{DM}_{i}&=&\tilde{\omega}_{a}\tilde{J}^{+}_{i}\tilde{J}^{-}_{i}+\tilde{\omega}_{c}\tilde{a}^{\dag}_{i}\tilde{a}_{i}+\beta(\tilde{a}_{i}\tilde{J}^{+}_{i}+\tilde{a}^{\dag}_{i}\tilde{J}^{-}_{i})
\end{eqnarray}
where $\tilde{\omega}_{a}=\omega_{a}-i\gamma_{a}$, $\tilde{\omega}_{c}=\omega_{c}-i\gamma_{c}$. $\gamma_{a}$ and $\gamma_{c}$ are decay rates of atoms and cavities, respectively. $\tilde{a}_{i}^{\dag}$($\tilde{a}_{i}$) is a quasiboson creation(annihilation) operator. $\tilde{J}^{+}_{i}$($\tilde{J}^{-}_{i}$) is the dressed atomics raising(lowering) angular momentum operator. The dissipation becomes an inherent property for the DBH model considered here.

A superfluid order parameter $\psi$, with the mean field assumption $\psi\equiv\langle \tilde{a}_{i}\rangle$, is usually introduced to gain insight into the role of dissipation in the quantum phase transition. For $\psi\neq0$, the system is in superfluid phase. When $\psi=0$, the system is in the Mott-insulator phase. In the present case, the expected value of $\tilde{a}_{i}$ is in general complex with the formation $\langle \tilde{a}_{i}\rangle=\psi-i\psi_{\gamma}$. $\psi_{\gamma}$ is a solvable small quantity
as a function of decay rates of the system, and vanishes in the limit of ideal cases. Using the decoupling approximation $\tilde{a}^{\dag}_{i}\tilde{a}_{j}=\langle\tilde{a}^{\dag}_{i}\rangle\tilde{a}_{j}+\langle\tilde{a}_{j}\rangle\tilde{a}^{\dag}_{i}-\langle\tilde{a}^{\dag}_{i}\rangle\langle\tilde{a}_{j}\rangle$. The mean-field Hamiltonian of Eq.(2.2) can be written as
\begin{eqnarray}
H^{MF} &=&\sum_{i}H_{i}^{MF}  \\
H^{MF}_{i} &=& H_{i}^{DM}-\kappa\psi(\tilde{a}^{\dag}_{i}+\tilde{a}_{i})+\kappa|\psi|^{2}-\mu\sum_{i}\tilde{N_{i}}\nonumber
\end{eqnarray}
This mean-field Hamiltonian is assumed to be the same for every site.

\section{Eigenvalues and eigenstates of the dissipative Dicke-Bose-Hubbard model}
In the following, the case of two TLAs in each cavity is investigated as an example to provide a detailed illustration. The extension for an arbitrary number of two-level atoms is given in \uppercase\expandafter{\romannumeral4}, which can be easily calculated by using the
same approach. The bare states of system are $|0,e^{\otimes2}\rangle|n-2\rangle$, $|g,e\rangle|n-1\rangle$, and $|g^{\otimes2},0\rangle|n\rangle$ with photon number $n$ running from $0$, $1$, $2$, $3$ to $\infty$\cite{Lei}. For two TLAs system, the case in which two atoms are in the excited state can be denoted as $|0,e^{\otimes2}\rangle$, only one atom in the excited state is denoted by $|g,e\rangle$, and $|g^{\otimes2},0\rangle$ is for the case that the two atoms are in the ground state. Here the total of $3n$ bare state bases form a group for the whole Hilbert space. Based on these states, the matrix elements for $H^{MF}_{n}$ can be obtained.
 \begin{widetext}
 \begin{eqnarray}
H^{MF}_{n} = \left|
\begin{array}{ccc}
2\tilde{\omega}_{a}+(n-2)\tilde{\omega}_{c}-n\mu  &   \sqrt{2(n-1)\beta}  &   0\\
\sqrt{2(n-1)\beta} &  2\tilde{\omega}_{a}+(n-1)\tilde{\omega}_{c}-(n+1)\mu  &  \sqrt{2n\beta}\\
0  &   \sqrt{2n\beta}   &   n\tilde{\omega}_{c}-n\mu  \\
\end{array}
\right|+\kappa|\psi|^{2}
\end{eqnarray}
\end{widetext}
with $\tilde{\omega}_{c}=\tilde{\omega}_{a}=\tilde{\omega}(\tilde{\omega}=\omega-i\gamma)$, $\gamma=\gamma_{a}+\gamma_{c}$. The eigenvalues can be obtained by diagonalizing the matrix in Eq.$(3.1)$, and the corresponding eigenstates can be found.
\begin{eqnarray}
E^{(0)}_{|0,n\rangle}=n\tilde{\omega}
\end{eqnarray}
\begin{eqnarray}
 E^{(0)}_{|\pm,n\rangle}&=&\frac{(2n+1)\tilde{\omega}\pm\beta R(n,\frac{\tilde{\omega}}{\beta})}{2}
\end{eqnarray}
\begin{widetext}
\begin{eqnarray}
|0,n\rangle=\frac{-\sqrt{n-1}|0,e^{\otimes2}\rangle|n-2\rangle+\sqrt{n}|g^{\otimes2},0\rangle|n\rangle}{\sqrt{2n-1}}
\end{eqnarray}
\begin{eqnarray}
|\pm,n\rangle&=&\frac{\sqrt{n}|0,e^{\otimes2}\rangle|n-2\rangle+\frac{1}{2\sqrt{2}}[\frac{\tilde{\omega}}{\beta}\pm R(n,\frac{\tilde{\omega}}{\beta})]|g,e\rangle|n-1\rangle+\sqrt{n-1}|g^{\otimes2},0\rangle|n\rangle}{\sqrt{2n-1+\{\frac{1}{2\sqrt{2}}[\frac{\tilde{\omega}}{\beta}\pm R(n,\frac{\tilde{\omega}}{\beta})]\}^{2}}}
\end{eqnarray}
\end{widetext}
Here $R(n,\frac{\tilde{\omega}}{\beta})=\sqrt{8(2n-1)+(\frac{\tilde{\omega}}{\beta})^{2}}$ is the effective Rabi frequency. The energy levels split into three branches corresponding to the upper branch $E^{(0)}_{|+,n\rangle}$, center branch $E^{(0)}_{|0,n\rangle}$ and the lower branch $E^{(0)}_{|-,n\rangle}$ as shown in Fig. $1(b)$.
\section{ the quantum phase transition}
In this section, we use the perturbation theory to obtain the superfluid order parameter and study the quantum phase transition by changing the controlling parameters. We have assumed that cavities are coupled weakly to each other, thus, the interaction term between cavities can be considered as a perturbation term when the two-level atoms system is coupled strongly to the cavity field. The effective Hamiltonian Eq.$(2.3)$ thus reads
\begin{eqnarray}
H^{MF}_{i}&=&H^{DM}_{i}+H^{\prime}_{i}\nonumber\\
H^{\prime}_{i}&=&-\kappa\psi(\tilde{a}^{\dag}_{i}+\tilde{a}_{i})+\chi|\psi|^{2}-\mu \tilde{N_{i}}
\end{eqnarray}
which is valid on each site, we therefore drop the subscript $i$ in the following.
Considering the analogy of transition from the Mott-insulator to superfluid state between the Jaynes-Cummings model and the Bose-Hubbard model and the fact that the analytical results obtained by the second and fourth-order perturbations are in good agreement with the exact diagonalization numerical calculation\cite{Oosten}, we derive the analytically solution of the system in terms of the second-order perturbation for simplicity.  Eq.$(3.2)$ and Fig. $1(b)$ show that a center energy level $E_{|0,n\rangle}$ is required to perform the translation, thus, the on-site repulsion $U$ based on the center branch state $|0,n\rangle$ is independent of the atom-cavity coupling $\beta$,  which is different from one defined by the state $|\pm,n\rangle$. To study the quantum phase transition in detail, the superfluid order parameter must calculated separately for different cases.

\emph{Preparing in the center branch of the dressed state}: According to the definition of the superfluid order parameter $\psi=\langle\Phi_{n}(t)|\tilde{a}_{i}|\Phi_{n}(t)\rangle$, $|\Phi_{n}(t)\rangle$ can be obtained based on the second-order perturbation theory. We first obtain the second-order corrections of energy eigenvalues $E^{(2)}_{|0,n\rangle}$ and (normalized) eigenstates $\tilde{\phi}^{(2)}_{|0,n\rangle}$ with respect to the dressed basis Eq.$(3.4)$.
\begin{eqnarray}
E^{(2)}_{|0,n\rangle}=\frac{(n-1)(2n-2)^{2}\kappa^{2}\psi^{2}}{(2n-1)(2n-3)(\epsilon-i\gamma)}\nonumber \\ +\frac{4n^{3}\kappa^{2}\psi^{2}}{(2n-1)(2n+1)(-\epsilon+i\gamma)} \end{eqnarray}
\begin{eqnarray}
\tilde{\phi}^{(2)}_{|0,n\rangle}=\frac{\sqrt{n-1}(2n-2)(-\kappa\psi)}{\sqrt{(2n-1)(2n-3)}(\epsilon-i\gamma)}|n-1\rangle\nonumber \\+\frac{2n\sqrt{n}(-\kappa\psi)}{\sqrt{(2n-1)(2n+1)}(\epsilon-i\gamma)}|n+1\rangle
\end{eqnarray}
where $\epsilon=\omega-\mu$. Therefore, the eigenvalue of the dissipative system based on the second-order perturbation theory is
\begin{eqnarray}
E_{|0,n\rangle}\equiv E_{s}+iE_{\gamma}
\end{eqnarray}
with
\begin{eqnarray}
E_{s}&=&n\epsilon+\kappa|\psi|^{2}+\frac{(-8n^{3}+12n^{2}+4n-4)\kappa^{2}\psi^{2}\epsilon}{(2n-1)(2n-3)(2n+1)(\epsilon^{2}+\gamma^{2})} \nonumber\\
E_{\gamma}&=&n\gamma+\frac{(-8n^{3}+12n^{2}+4n-4)i\kappa^{2}\psi^{2}\gamma}{(2n-1)(2n-3)(2n+1)(\epsilon^{2}+\gamma^{2})} \nonumber
\end{eqnarray}
When the system is in the Mott-insulator state, $\psi=0$, we have $E_{\gamma}= n\gamma$. When $\psi\neq 0$, one can take $E_{\gamma}\approx n\gamma$ because we assume that the coupling strength $\kappa$ between cavities is weak.  Up to second order, the expression for the (normalized) eigenstates is
\begin{eqnarray}
\phi_{|0,n\rangle}=\frac{1}{\sqrt{\tilde{N}}}\tilde{\phi}_{|0,n\rangle}
\end{eqnarray}
\begin{eqnarray}
\tilde{\phi}_{|0,n\rangle}=\frac{\sqrt{n-1}(2n-2)(-\kappa\psi)}{\sqrt{(2n-1)(2n-3)}(\epsilon-i\gamma)}|n-1\rangle+|n\rangle\nonumber
 \\+\frac{2n\sqrt{n}(-\kappa\psi)}{\sqrt{(2n-1)(2n+1)}(\epsilon-i\gamma)}|n+1\rangle \nonumber
\end{eqnarray}
\begin{eqnarray}
\tilde{N}=1+\frac{(n-1)(2n-2)^{2}\kappa^{2}\psi^{2}}{(2n-1)(2n-3)(\epsilon^{2}+\gamma^{2})}\nonumber \\+\frac{4n^{3}\kappa^{2}\psi^{2}}{(2n-1)(2n+1)(\epsilon^{2}+\gamma^{2})}\nonumber
\end{eqnarray}
$\tilde{N}$ is the normalized constant. For the open system considered here, the superfluid order parameter $\psi$ is time-dependent. According to Eq.$(4.5)$, the (normalized) eigenstates is a function of the time, however, its time derivative can be ignored because of the second-order correction. Thus, the approximative time-dependent wave function of the system can be written as
\begin{eqnarray}
\Phi_{n}(t)=f(t)\phi_{|0,n\rangle}\nonumber
\end{eqnarray}
Using the Schr\"{o}dinger equation, one can find
\begin{eqnarray}
\Phi_{n}(t)=\phi_{|0,n\rangle}e^{-iE_{|0,n\rangle}t}
\end{eqnarray}
Therefore, the superfluid order parameter $\psi$ for the state $|0,n\rangle$ can be obtained
\begin{widetext}
\begin{eqnarray}
\psi_1=e^{-n\gamma t}\sqrt{\frac{(8n^{3}-12n^{2}-4n+4)\epsilon}{(16n^{4}-32n^{3}+12n^{2}+4n-4)\kappa}-\frac{(2n-1)(2n+1)(2n-3)(\epsilon^{2}+\gamma^{2})}{(16n^{4}-32n^{3}+12n^{2}+4n-4)\kappa^{2}e^{-2n\gamma t}}}\label{one}
\end{eqnarray}\\
\end{widetext}
Eq. $(4.7)$ shows that $\psi_{1}$ is a function of the parameters $\kappa$, $\gamma$, $t$ and $\mu$ (In present case, $\mu$ is a constant). The superfluid order parameter evolves and decays with time with a decay rate proportional to the number of photons $n$.

\emph{Preparing in the negative branch of the dressed state}: Assume that each site is prepared in the
negative branch of the dressed state $|-,n\rangle$. We can find the second-order deviations using a similar procedure, although the calculations become quite tedious when using our current formulation. The superfluid order parameter $\psi_{2}$ can be obtained by solving the following equation.
\begin{widetext}
\begin{eqnarray}
\psi_2=Re\{\frac{e^{-2\gamma nt}}{\tilde{N^{'}}}[\frac{2[2\sqrt{n(n-1)(n-2)}+\frac{\sqrt{n-1}}{8}(\frac{\omega+i\gamma}{\beta}-R_{n-1}^{\dag})(\frac{\omega-i\gamma}{\beta}-R_{n})]^{2}(-\kappa\psi_{2})}
{[2\epsilon+2i\gamma-\beta(R^{\dag}_{n}-R_{n-1}^{\dag})][2n-1+\frac{1}{8}(\frac{\omega-i\gamma}{\beta}-R_{n-1})^{2}][2n-3+\frac{1}{8}(\frac{\omega+i\gamma}{\beta}-R^{\dag}_{n-1})^{2}]}+\\ \nonumber
\frac{2[2\sqrt{n(n-1)(n+1)}+\frac{\sqrt{n}}{8}(\frac{\omega+i\gamma}{\beta}-R_{n}^{\dag})(\frac{\omega-i\gamma}{\beta}-R_{n+1})]^{2}(-\kappa\psi_{2})}
{[-2\epsilon+2i\gamma-\beta(R_{n}-R_{n+1})][2n-1+\frac{1}{8}(\frac{\omega+i\gamma}{\beta}-R^{\dag}_{n})^{2}][2n+1+\frac{1}{8}(\frac{\omega-i\gamma}{\beta}-R_{n+1})^{2}]}]\}\label{two}
\end{eqnarray}
\end{widetext}
$\tilde{N^{'}}$ (Appendix) is the normalized constant. In what follows, we use Eqs. $(4.7)$ and $(4.8)$  to numerically investigate the features of quantum phase transition arising from the competition between the on-site repulsion $U_{n}$ and the hopping rate under the influences of the environment.
\begin{figure}
\includegraphics[width=5.5cm]{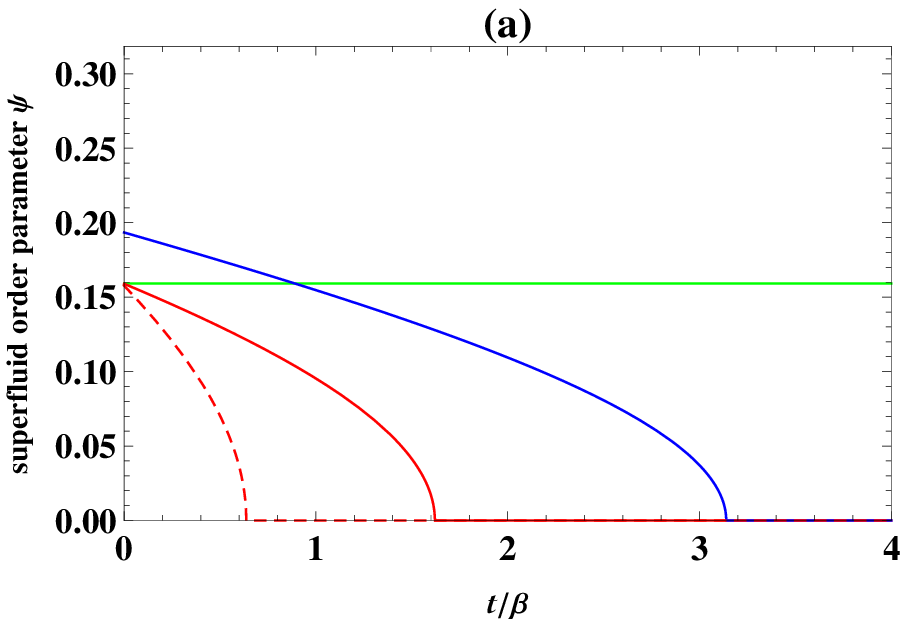}
\includegraphics[width=5.5cm]{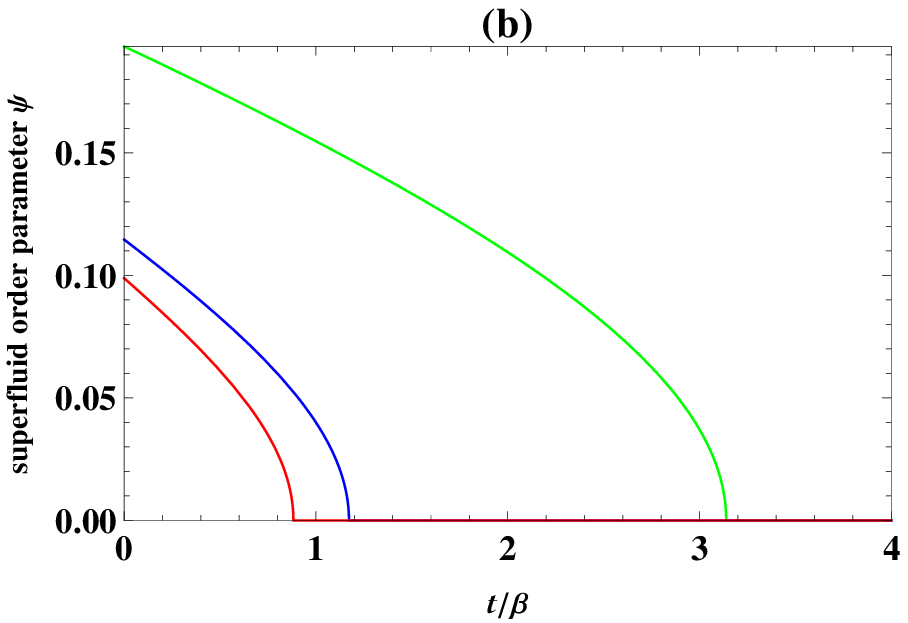}
\includegraphics[width=5.5cm]{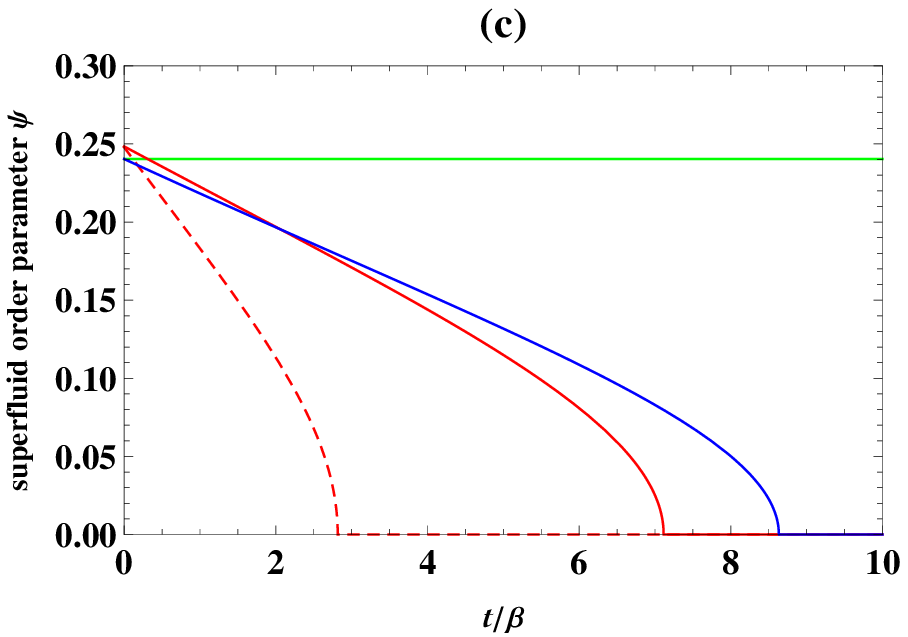}
\includegraphics[width=5.5cm]{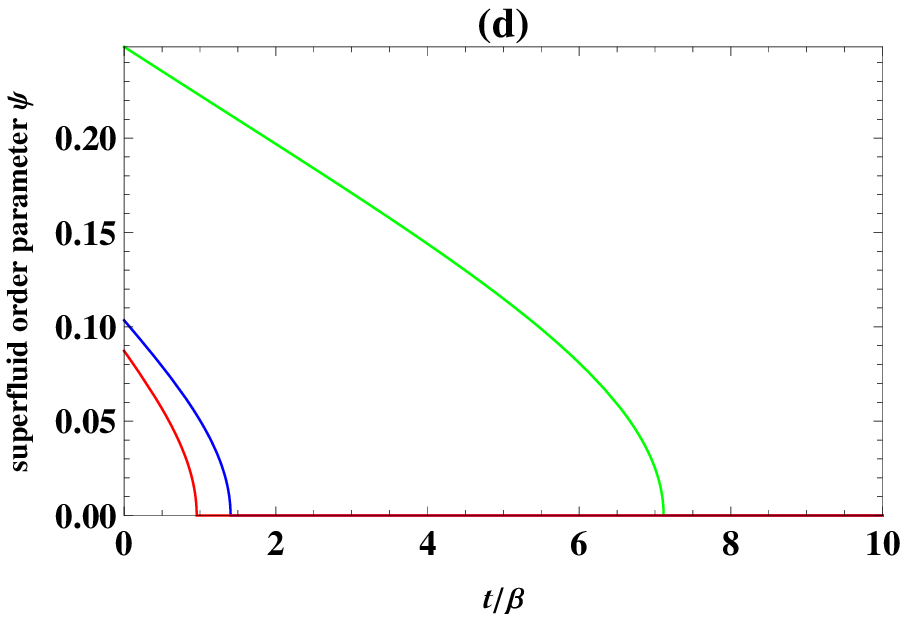}
\caption{(Color online) The decrease of the superfluidity on each site for a initial superfluid state. (a) and (b) are for the state $|0,n\rangle$, but (c) and (d) are for the state $|-,n\rangle$. The resonance frequency $\omega=10$, $\epsilon=0.7836$. The superfluid order parameter decays continuously and beyond $t_{c}$ the system behaves as a Mott-insulator-like states. In (a) and (c), the parameters are $n=3$,  $(\gamma/\beta,\kappa/\beta)=(0,1)$(Green), $(0.02,1)$(Red), $(0.05,1)$(Dashed red), and $(0.02,1.2)$ (Blue). In (b) and (d) $\gamma/\beta=0.02$, $\kappa/\beta=1.2$, with different $n$: $n=3$ (Green), $n=9$ (Blue) and $n=12$ (Red).  The long-range order decays rapidly when $n$ increases.}
\end{figure}
\begin{figure}
\includegraphics[width=5.5cm]{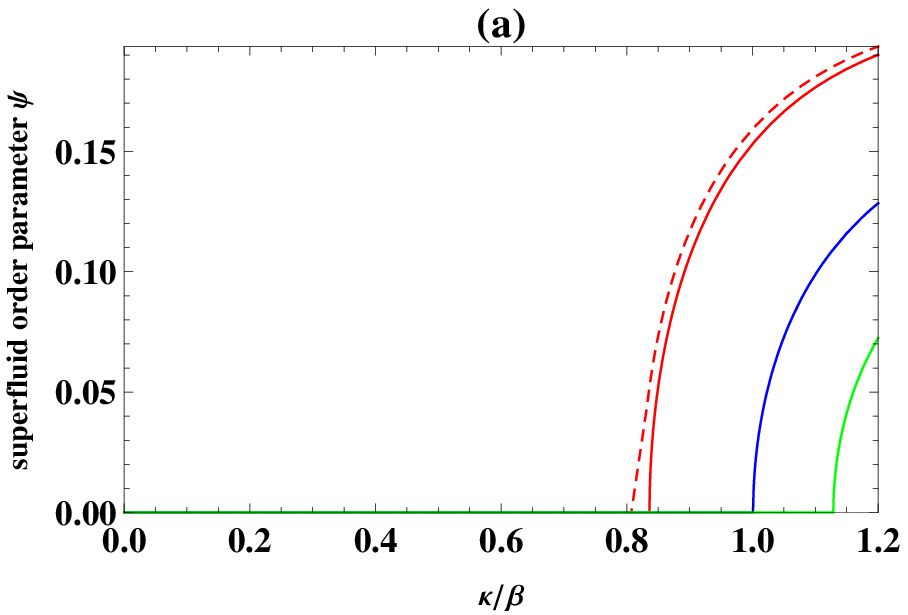}
\includegraphics[width=5.5cm]{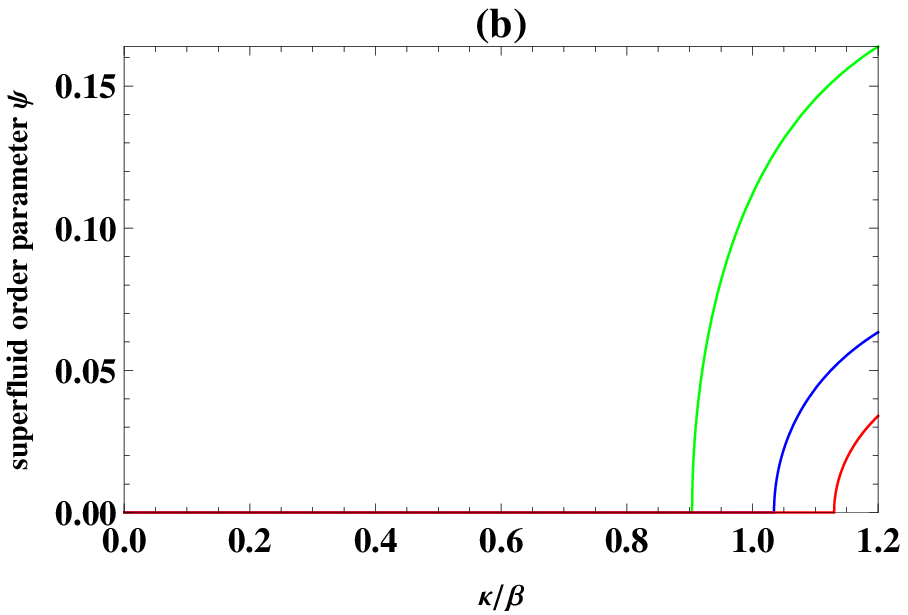}
\includegraphics[width=5.5cm]{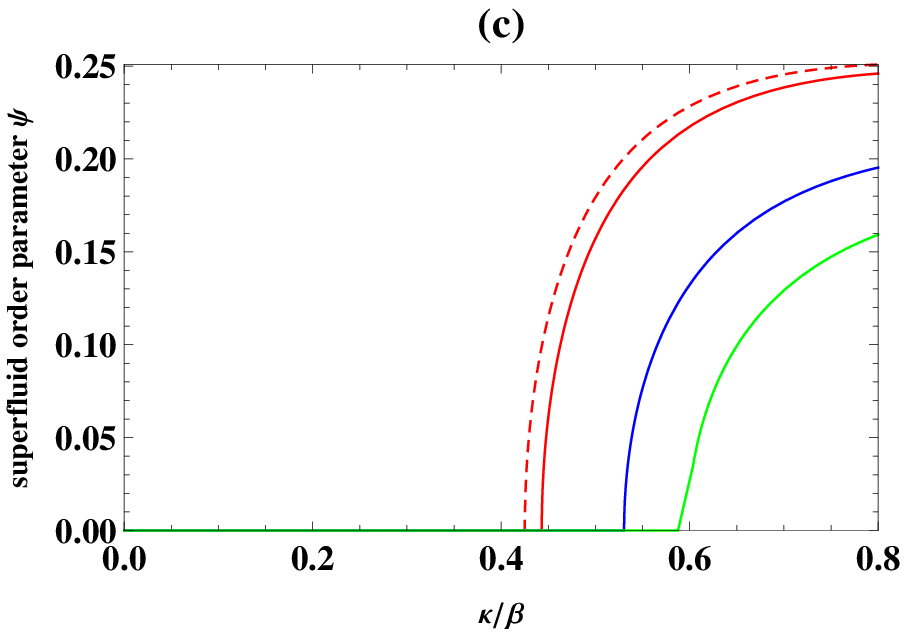}
\includegraphics[width=5.5cm]{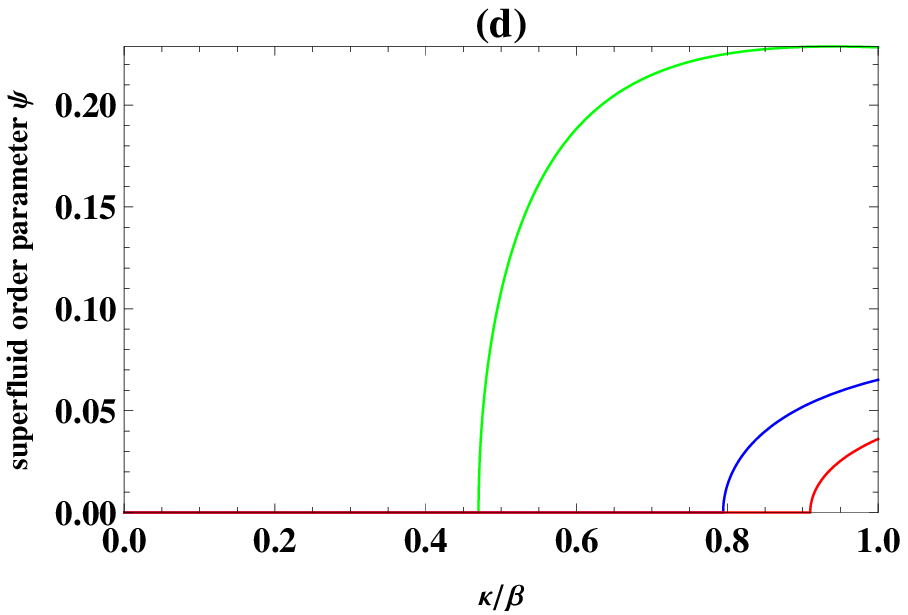}
\caption{(Color online) The restoring of long-range coherence from the Mott-insulator state. (a) and (b) are for the state $|0,n\rangle$, (c) and (d) are for the state $|-,n\rangle$. The system can reach a superfluid phase with  continuously increasing the intercavity coupled rate $\kappa$. In (a) and (c), the parameters are $n=3$,  $(\gamma/\beta,t)=(0,0)$(Dashed), $(0.05,0)$(Red), $(0.05,0.3)$(Blue), and $(0.05,0.5)$ (Green). In (b) and (d) $\gamma/\beta=0.05$, $t=0.3$, with different $n$: $n=3$ (Green), $n=9$ (Blue) and $n=12$ (Red). With an increase of $n$, a larger value of the intercavity hopping is needed.}
\end{figure}

\emph{Analyses}. As illustrated in Fig. $2$. First, we start with a superfluid phase. The time evolution of the superfluid order parameter $\psi$ for different hopping rates $\kappa$ and decay rates $\gamma$ are shown. Comparing Figs. $2a(b)$ with $2c(d)$, a clear quantum phase transition is found for different initial states. The ideal cases are also given in Fig. $2$ for comparison, which shows that the system is still in the coherent state that was prepared initially.  The evolution of the dissipative system clearly reflects the expected decay of the coherence, which is the most obvious characteristic different from ideal cases. For a small $t$, although $\psi$ decreases slightly, the system remains in a superfluid state. At a sufficiently large $t$, the effects of the environment become large, and the coherence of the system is initially destroyed in a pronounced manner and is then gradually reduced. Thus, $\psi$ decays rapidly and the system undergoes a phase transition into a Mott-insulating phase. With the increase of the photon number $n$, the long-range order parameter will decrease rapidly, as shown in Fig. $3(b)$ and $(d)$, because of the decay time is  proportional to $n$.  The critical point $t_c$ is a function of controlling parameters and can be found by setting $\psi=0$ in Eqs. $(4.7)$ and $(4.8)$, which yields $t_{c}=\frac{1}{2n\gamma}\ln\frac{(4n^{2}-4n-4)\kappa}{(2n+1)(2n-3)\epsilon}$. It follows that, for certain cavity decay rates $\kappa$, one may change the other controlling parameters according to $t_c$ to enable the dissipative system to maintain coherence for a relatively long time.  Obviously, when the
external environment is considered, the decoherence
of every resonator and the TLAs would result in the decay of the superfluid order parameter. In the experiment, dynamical decoupling\cite{Bylander} and feedback control\cite{Xue} have been proposed to hamper the decay of the cavity field and TLAs and thus improved the coherence time.
\begin{widetext}
\begin{figure}
\includegraphics[width=9cm,height=9.5cm]{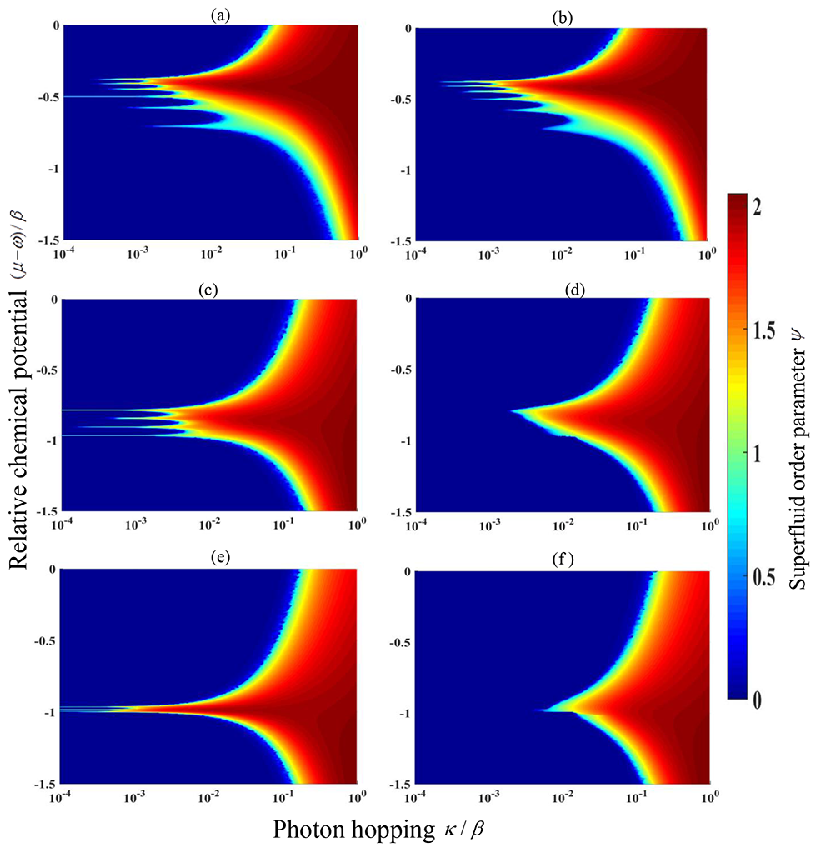}
\caption{(Color online) The superfluid order parameters as a function of the relative chemical potential and the hopping rates for arbitrary number of TLAs, (a) $2$ TLAs, (b) $2$ TLAs, (c) $4$ TLAs, (d) $4$ TLAs, (e) $6$ TLAs, (f) $6$ TLAs. (a), (c) and (e)are for the ideal cases, others are for the dissipative cases. We have chosen  $n=8$ and  $\gamma=0.2$. The change of Mott-insulator and superfluid phase boundary can be seen clearly.}
\end{figure}
\end{widetext}

In contrast, as shown in Fig. $3$, we seek to determine how a Mott-insulator state in the beginning can restore the coherence by changing the intercavity hopping rate $\kappa$ for a dissipative system.
For a small $\kappa$, there are no enough excitations for hopping between cavities. By rasing the hopping rate to a certain value $\kappa_{c}=\frac{(8n^{3}-12n^{2}-2n+3)\epsilon e^{2n\gamma t}}{(8n^{3}-12n^{2}-4n+4)}$, the system will restore its long-range coherence and a phase transition from Mott-insulator to the superfluid phase appears. According to Eqs.$(4.7)$ and $(4.8)$, the photon hopping rate is also found to decrease because of the effect of the environment, thus the long-range coherence can only occur when the increase of the photon hopping rate is faster than its decay. Figs. $3(a)$ and $3(c)$ also demonstrate that the influence of environment accumulate over time. With an increase in time, a large hopping rate is required to restore the coherence. Because the system and the environment have been recognized as a whole system in the effective Hamiltonian, then the dissipation is the inherent nature of the system. Therefore, for $t=0$, the system is also dissipative, and the hopping rate required for the phase transition to occur is higher than the rate expected in the ideal case.  In addition, increasing the number of photons to $n$, the dissipation of the system is also enhanced correspondingly, a higher hopping energy is thus required to induce a phase transition, as shown in Figs. $3(b)$ and $3(d)$.

In what follows, we extend the model to an arbitrary number of TLAs cases. The Dressed-state basis can be written by the general method to  diagonalize the effective DBH Hamiltonian (1.1) by numerical computation. The phase diagrams of the dissipative DBH model are plotted in Fig.$4$. For comparison, we also show the ideal cases. In dissipative cases, we choose $t=0$, which implies that the dissipative system is nearly equilibrium. As shown in Fig. $4$, as interaction with the environment destroys the coherence of the system, the Mott lobes becomes smaller and the area of the coherent phase decreases. Next, the realization of the superfluid state requires a large hopping rate to derive the localized photons in each cavity. It can also be found that, in a regime with a small hopping rate $\kappa$, fewer TLAs could cause the system to become a localized phase compared with the ideal cases. With an increase in the number of TLAs, the coherent state may disappear rapidly for the dissipative system.

\section{Conclusion}
Based on the quasibosonic approach, a realistic situation of a DBH  model coupled to its environment was considered. The analytical solution of the superfluid order parameter for two TLAs per cavity was derived. The transition behaviors of the superfluid to Mott-insulation phase and the restoring coherence were discussed. The phase diagram for an arbitrary number of TLAs was also investigated. As the number of TLAs increases, Mott lobes may disappear and such a system tends to be classical. Most importantly, the atomic number is far lower than that under ideal circumstances. This work can provide parameters for reference to simulate strongly correlated many body systems in the actual operation.

\begin{acknowledgments}
This work was supported by the National Natural Science
Foundation of China under Grant No.11274148.
\end{acknowledgments}

\section{Appendix}
The normalized constant of superfluid order parameter $\psi_{2}$.
\begin{widetext}
\begin{eqnarray}
\tilde{N}^{'}&=&4\kappa^{2}\psi^{2}AA^{\dag}+4\kappa^{2}\psi^{2}BB^{\dag} \nonumber
\end{eqnarray}\\
\begin{eqnarray}
A=\frac{2\sqrt{n(n-1)(n-2)}+\frac{\sqrt{n-1}}{8}(\frac{\omega+i\gamma}{\beta}-R_{n}^{\dag})(\frac{\omega-i\gamma}{\beta}-R_{n-1})}
{[2\epsilon-2i\gamma-\beta(R_{n}-R_{n-1})]\sqrt{[2n-1+\frac{1}{8}(\frac{\omega+i\gamma}{\beta}-R^{\dag}_{n})^{2}][2n-3+\frac{1}{8}(\frac{\omega-i\gamma}{\beta}-R_{n-1})^{2}]}}\nonumber
\end{eqnarray}\\
\begin{eqnarray}
B=\frac{2\sqrt{n(n-1)(n+1)}+\frac{\sqrt{n}}{8}(\frac{\omega+i\gamma}{\beta}-R_{n}^{\dag})(\frac{\omega-i\gamma}{\beta}-R_{n+1})}
{[-2\epsilon+2i\gamma-\beta(R_{n}-R_{n+1})]\sqrt{[2n-1+\frac{1}{8}(\frac{\omega+i\gamma}{\beta}-R^{\dag}_{n})^{2}][2n+1+\frac{1}{8}(\frac{\omega-i\gamma}{\beta}-R_{n+1})^{2}]}}\nonumber
\end{eqnarray}\\
\end{widetext}
 The conjugates of $A$ and $B$ are $A^{\dag}$ and $B^{\dag}$, respectively.
\end{document}